\documentclass[conference]{IEEEtran}
\IEEEoverridecommandlockouts
\usepackage{cite}
\usepackage{amsmath,amssymb,amsfonts}
\usepackage{graphicx, caption, subcaption}
\usepackage{textcomp}
\usepackage[table]{xcolor}
\usepackage{algorithm,algpseudocode}
\usepackage{hyperref}
\hypersetup{
     colorlinks = true,
     citecolor = green,
     }

\usepackage{multirow}
\usepackage{booktabs}
\usepackage[usestackEOL]{stackengine}

\def\BibTeX{{\rm B\kern-.05em{\sc i\kern-.025em b}\kern-.08em
    T\kern-.1667em\lower.7ex\hbox{E}\kern-.125emX}}
\begin{document}


\newboolean{showcomments}
\setboolean{showcomments}{true} 
\ifthenelse{\boolean{showcomments}}
  {
		\newcommand{\nbb}[2]{
		\fcolorbox{black}{yellow}{\bfseries\sffamily\scriptsize#1}
		{\sf$\blacktriangleright$\textcolor{blue}{\textit{#2}}$\blacktriangleleft$}
		}
		\newcommand{\version}{\emph{\scriptsize$-$9.2.2011$-$}}
		\newcommand{\remarks}[1]{\color{red}[#1]\color{black}}
		\newcommand{\copied}[1]{\color{green}[#1]\color{black}}
		\newcommand{\modified}[1]{\color{blue}[#1]\color{black}}
		\newcommand{\raw}{$\rightarrow$}
		\newcommand{\ins}[1]{\textcolor{blue}{\uline{#1}}} 
		\newcommand{\del}[1]{\textcolor{red}{\sout{#1}}} 
		\newcommand{\chg}[2]{\textcolor{red}{\sout{#1}}{\raw}\textcolor{blue}{\uline{#2}}} 
		\newcommand{\ugh}[1]{\textcolor{red}{\uwave{#1}}} 
  }
  {
		\newcommand{\nbb}[2]{}
		\newcommand{\remarks}[1]{}
		\newcommand{\modified}[1]{#1}
		\newcommand{\copied}[1]{#1}
		\newcommand{\version}{}
		\newcommand{\ugh}[1]{#1} 
		\newcommand{\ins}[1]{#1} 
		\newcommand{\del}[1]{} 
		\newcommand{\chg}[2]{#2} 
  }

\newcommand{\jens}[1]{\nbb{Jens}{#1}}
\newcommand{\cbe}[1]{\nbb{CBe}{#1}}
\newcommand{\lars}[1]{\nbb{Lars}{#1}}
\newcommand{\sankar}[1]{\nbb{Sankar}{#1}}
\newcommand{\ce}[1]{\nbb{CE}{#1}}
\newcommand{\mb}[1]{\nbb{Markus}{#1}}
\newcommand{\comment}[1]{\nbb{Comment}{#1}}

\title{Making Lab Sessions Mandatory -- On Student Work Distribution in a Gamified Project Course on Market-Driven Software Engineering}


\author{
\IEEEauthorblockN{Markus Borg\\markus.borg@cs.lth.se}
\IEEEauthorblockA{Lund University\\Lund, Sweden}}

\maketitle

\begin{abstract}
Unfair work distribution in student teams is a common issue in project-based learning. One contributing factor is that students are differently skilled developers. In a course with group work intertwining engineering and business aspects, we designed an intervention to help novice programmers, i.e., we introduced mandatory programming lab sessions. However, the intervention did not affect the work distribution, showing that more is needed to balance the workload. Contrary to our goal, the intervention was very well received among experienced students, but unpopular with students weak at programming.
\end{abstract}


\section{Introduction}
Project courses that involve group work is an important part of Software Engineering (SE) education.
Project-based learning in group settings combines ``knowing'' and ``doing'' as students together solve relevant problems~\cite{markam_project_2011} and the positive effects on the learning experience of students are well-known in higher education.
By working in groups, the learning experience goes beyond the fundamental technical curriculum as students also practice ``soft skills'', e.g., leadership, communication, and planning -- aspects often reported to be lacking in engineering education~\cite{moreno_balancing_2012,marques_enhancing_2018}.

However, designing group work in education brings several challenges.
Simply providing student groups with project assignments does not automatically create the right conditions for successful project-based learning~\cite{markam_project_2011}.
Not only do poorly designed project assignments limit the potential for learning, but it also increases the risk of well-known challenges inherent in group work, e.g., student conflicts~\cite{borg_conflict_2011} and unfair work allocation.
The issue of unfair work allocation spans from freeriding and social loafing~\cite{duim_good_2007} to the other extreme such as domineering students~\cite{oakley_turning_2004}.

Designing group work in SE courses that encourages a balanced workload among students is hard.
Two related challenges are incompatible expectations and different software development skills.
Students often have considerable differences in motivation, i.e., different levels of ambition and commitment to the project tasks~\cite{cronholm_project_2006}. If the projects are graded, some students will aim for a passing grade while others will target top grades. Poor expectation alignment also increases the risk of conflicts in student groups~\cite{borg_conflict_2011}.
Students also have different backgrounds in terms of programming maturity and experience with the SE tool chain~\cite{suhonen_applications_2007}.

In this paper, we report on how we addressed both different motivation levels and different software development skills when evolving an SE project course at Lund University\footnote{The course material is available on https://github.com/lunduniversity/introsofteng under a Creative Commons license.}. 
To motivate students with different interests, we designed the project assignment to include gamification elements related to both engineering and business -- at the same time responding to calls for increased emphasis on business in SE education~\cite{moreno_balancing_2012}. Course evaluations show that students appreciate the gamification elements and indicate high engagement~\cite{kearsley_engagement_1998}. In relation to tackling the differences in software development skills, we present empirical results from before and after making practical lab sessions mandatory -- referred to as the \textit{intervention}.

The main contribution of this paper is an analysis, based on method triangulation, of the impact of making four lab sessions mandatory, starting in the first week of the course. A second contribution is a detailed presentation of our efforts to introduce gamification in a market-driven SE setup, i.e., the case description.
Three research questions guide our empirical investigation on mandatory lab sessions:
\begin{itemize}
\item[RQ1] Do mandatory labs help students distribute the project effort more evenly over the weeks?
\item[RQ2] Do mandatory labs balance work distribution within groups?
\item[RQ3] Is the intervention well-received by the students?
\end{itemize}

The paper is structured as follows. Section~\ref{sec:rw} introduces related work on group projects in SE education. Section~\ref{sec:bg} contains the case description, i.e., the SE course under study. In Section~\ref{sec:method}, we explain how we collect and analyze empirical data. Section~\ref{sec:res} presents our results and the primary threats to validity are elaborated in Section~\ref{sec:threats}. Finally, Section~\ref{sec:conc} reports our conclusion.

\section{Related Work} \label{sec:rw}
Several papers proposed what SE project courses should include and sometimes how they should be taught. Already in 2001, Meyer wrote that group work is fundamental to teach SE, but they are often insufficient to prepare students for real challenges, e.g., due to lack of end-user interaction~\cite{meyer_software_2001}. Ghezzi and Mandrioli argued that just studying textbooks and doing exercises is not real learning without a practical experience~\cite{ghezzi_challenges_2005}. They called for innovative project formats to create realistic projects. For example, partner companies can be involved to increase realism~\cite{gnatz_practical_2003}. We instead stimulate realistic end-user interaction by setting up an internal ecosystem in which students market their software. 

Designing an inner market for the course responds to several calls for the integration of business perspectives in SE education. Ghezzi and Mandrioli stated that students shall learn to understand the context of software, including the  ``social physical world'' of economics and business organizations~\cite{ghezzi_challenges_2005}, but they claimed that the teaching of such topics is too superficial and descriptive. Buffardi \textit{et al.} wrote that ``a lack of external pressures to make a product for real users'' obstructs realistic experiences~\cite{buffardi_tech_2017}. Mich hypothesized that SE courses lack business aspects as teachers ``consider functional specification and software development as a kind of `real' computer science, under-estimating business requirements analysis''~\cite{mich_teaching_2014}. 

Several papers suggested solutions to intertwine engineering and business. One solution is to use case competitions, as presented by Burge and Troy~\cite{burge_rising_2006}. They reported having partner companies providing business-oriented SE cases, and using the partners in a final jury. J\"arvi \textit{et al.} designed a course on ``Lean Software Startup'' mixing SE and business students~\cite{jarvi_lean_2015}. Similarly, Buffardi \textit{et al.} designed a course with mixed students from two programs, computer science and entrepreneurship, to create a setting of real software product development~\cite{buffardi_tech_2017}. Apart from entrepreneurship, previous work has also introduced the importance of business aspects from the perspective of requirements engineering, e.g., in relation to business process models~\cite{sedelmaier_using_2014} and requirements analysis~\cite{mich_teaching_2014}.



Students following our SE course are differently skilled programmers. Many students have a hard time during introductory programming courses~\cite{thomas_learning_2002,teague_collaborative_2008}, which might result in gaps in knowledge and skills. Unfortunately, the gaps might be further amplified in subsequent courses with collaboration elements. In the context of pair programming, Bowman \textit{et al.} found that students working with more experienced partners had substantially lower learning outcomes~\cite{bowman_how_2019}.

Several papers discuss challenges when teaching students with diverse skills and interests. Pieterse and Thompson introduced the term ``academic alignment'' to describe a group of students who are homogenous in terms of academic abilities, skills and goals~\cite{pieterse_academic_2010}. Weak alignment can cause group conflicts~\cite{borg_conflict_2011} and unbalanced skills within student teams have been reported as a problem when teaching agile software development~\cite{mckinney_affective_2005}. Forte and Guzdial shared negative experience from mixing computer science majors and non-majors in an introductory computing course, and recommend tailoring different courses for different student backgrounds~\cite{forte_motivation_2005}. Students equivalent to non-majors follow our course, i.e., the student cohort is diverse both in terms of skills and interests.

In some groups, possibly due to weak academic alignment, students' contributions will vary considerably. Social loafing or free-riding is frequently reported in the literature~\cite{duim_good_2007,borrego_team_2013}, and also the opposite referred to as ``diligent isolates'' by Pieterse and Thompson~\cite{pieterse_academic_2010}. To support balanced workload withing groups, we leave it to the students to form (academically aligned) groups and then we require everyone to sign group collaboration contracts~\cite{lee_collaboration_2015}. We argue that the literature on SE education has focused more on the assessment of individual contributions in group work, e.g.,~\cite{clark_self_2005}, than on approaches to support fair work distribution. In this paper, we investigate how making lab sessions mandatory influenced the distribution of student effort.

\section{Background and Case Description} \label{sec:bg}
This section describes the evolution of the course, its educational context, basic concepts of Robocode, and details about the course editions under study. The course material is actively maintained on GitHub and further informmtion can be found in the project instructions~\cite{github}.

\subsection{Formalities and Evolution of the Course}
Public higher education in the country is grant-aided and free of charge. The academic year is divided into two semesters, and the course under study runs in the late spring. The course is not open to international students and is taught in Swedish.

The course corresponds to 6 ECTS credits, i.e., 4 weeks of full-time study. The course prerequisites are limited to having completed labs and hand-in exams in an introductory programming course, i.e., passing the final programming exam is not required. Apart from the project task, the course consists of 7 lectures, 4 classroom exercise sessions, and 4 computer lab sessions~\cite{github} -- each element is scheduled for 90 min. All exercise and lab sessions are highly relevant to the projects.

Serving as a gateway to more advanced elective SE courses, the curriculum includes brief introductions to development processes, requirements engineering, software design, and software testing. Students practice programming in an IDE, git, test automation, writing technical documentation, formal inspections, developing a business plan, and working in a team. 

The origins of the course can be traced back at least two decades and several different teachers have been involved. In the early days, the project assignment was limited to writing a project plan, a requirements specification, and a test specification and handing them in according to a waterfall process -- a format in line with recommendations in the 1990s~\cite{budgen_teaching_1991}. In a major revision of the course, the project task was replaced with waterfall development of a small garage management system with simulated hardware.

In 2017, the waterfall process model used by the student groups was replaced by the Unified Process to move away from the previous linear sequential waterfall flow. The year after, another major revision of the course took place and the garage project task was replaced by the current Robocode project. At the same time, an iterative process model was introduced consisting of three development sprints. Finally, before the 2019 edition of the course, we introduced the intervention discussed in this paper.

\subsection{Student Cohort} \label{sec:cohort}
The student cohort's diversity is inherently challenging with roughly the same number of students from two engineering programs. The course is compulsory for the study program Information and Communication Technology (ICT) in the second semester and elective compulsory for the Engineering Management (MGM) program, i.e.,  compulsory for third year MGM students that select an ICT profile. The MGM students have completed two programming courses before, but the ICT students only one. Furthermore, the entry requirements for the ICT program are low whereas the opposite applies to the MGM program.


From the teacher's perspective, there are three main differences between the student profiles. First, despite following a generalist engineering program, \textit{the MGM students are better software developers}. Both ICT and MGM students know fundamental Java constructs, but the latter know also basic algorithms and data structures, as well as how to execute JUnit test suites. Second, \textit{the MGM students are considerably more mature}. Two additional years at the campus means established peer networks and study techniques -- and more experience of project tasks and written deliverables. Third, \textit{the MGM students tend to aspire for higher grades} than the ICT students.

\subsection{Fundamentals of Robocode} \label{sec:robocode}
The goal of Robocode is to implement the behavior of a robot to compete against other robots in a battle arena. The contestants have no direct influence during the battles in the game, instead they develop the AI of the robot telling it how to behave and react to events. Robocode battles, referred to as ``rumbles'', are running in real-time and on-screen. Rumbles are either in the form of duels between individual robots, free-for-all battles with multiple robots, or battles between robot teams with up to 10 robots.

The objective of a rumble is to reduce the energy of enemy robots to 0. A standard robot starts with 100 energy and is equipped with a radar and a gun turret. Each time the gun is fired, the robot loses energy according to the payload. If the bullet does not hit any target, the energy is lost. If the bullet hits another robot, energy is transferred from the hit robot to the firing robot. Robot developers mainly write source code to 1) sweep the radar over the battlefield to detect enemies, 2) fire toward the most probable future enemy positions, 3) move the ego-robot to avoid being hit, and 4) communicate with other robots in the team\footnote{To support robot communication, we have standardized a protocol: RoboTalk~\cite{github}.}.

Robocode has been successfully used in educational contexts before, both for teaching Java programming~\cite{long_just_2007} and SE. Johnson designed an SE course where students developed robots while at the same time learning several techniques and tools~\cite{johnson_robocode_2011}. We reused some of the work by Johnson, such as the RobotTestbed to support automated system testing of robots. However, group work was not in focus in Johnson's course. The course design most similar to what we present in this paper was developed by Georgas~\cite{georgas_teams_2011}, teaching student teams about requirements gathering, software design, and technical writing. However, we present a novel approach to use Robocode to teach market-driven SE by introducing the concept of a robot market.



\subsection{Project Overview and Gamification}
The projects are done by groups of six students, i.e., reasonably small groups but enough to illustrate important aspects of collaborative work. Before the first lecture, students receive an email requesting them to prepare a mandatory group forming exercise at the first lecture. During that lecture, group sign-up sheets are prepared and students put their names in groups on a first-come, first-served basis. While different approaches to establishing student groups are possible, letting the students find peers to team up with reduces initial tension and tends to pool students that are academically aligned. Be it good or bad, there are typically no groups mixing ICT and MGM students, except a few exceptions when the number of students was not evenly divisible by six. 

Each group must develop a robot while adhering to a specified SE process (see Section~\ref{sec:process}) including practices such as product prototyping, integrated requirements engineering, and test automation. Furthermore, each group will compose a robot team to compete in a live rumble at the final lecture. However, no group is allowed to field their own robot -- instead, robots developed by other groups must be purchased on an open market. Consequently, each project group has three primary goals: 1) \textit{developing a successful robot} 2) \textit{maximizing profit} by selling many robots on the market, and 3) \textit{winning the rumble} by purchasing robots from other groups to compose a competitive Robocode team. Groups pursue these goals by completing three main activities that we refer to as \textit{engineering}, \textit{monetizing}, and \textit{strategizing}, respectively.

\begin{figure}
\centering
\includegraphics[width=0.48\textwidth]{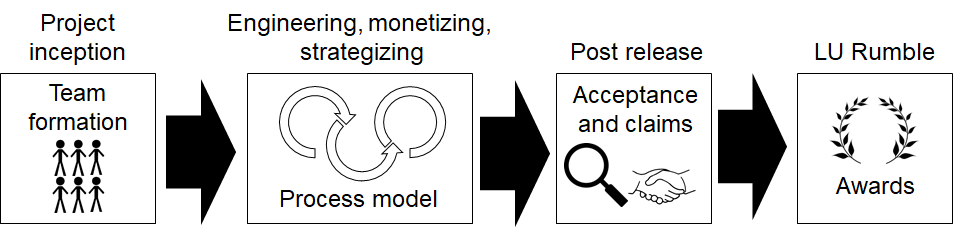}
\caption{The four phases of the project.}
\label{fig:overview}
\end{figure}

Fig.~\ref{fig:overview} shows the four main phases of the project. First, during the course inception, groups are formed and we introduce the project task. Second, the backbone of the course follows: three development sprints mixing engineering, monetizing, and strategizing. Third, groups perform acceptance testing of the robot they purchased to ensure that the delivered robot fulfills the expectations -- otherwise, purchasers may file business claims to request a cost deduction. Fourth, the rumble takes place followed by an awards ceremony with leaderboards. Three awards are given~\cite[Sec 3.4]{github}:

\begin{itemize}
\item The \textbf{Strategizer Award} goes to the winners of the rumble.
\item The \textbf{Monetizer Award} goes to the most profitable group, i.e., the most successful group on the market.
\item The \textbf{Engineering Award} goes to the Most Valuable Robot (MVR) during the rumble.
\end{itemize}

Organizationally, each group has three divisions as shown in the lower part of Fig.~\ref{fig:ecosystem}: 1) Engineering, 2) Executive, and 3) Sales. Everyone in the group will shoulder several roles, as each student will be involved in the Engineering division as developers and testers -- supported by the mandatory lab sessions. Fig.~\ref{fig:ecosystem} indicates this by the six small figures denoted ``Dev/Test''. On top of this,  groups also assign the following additional roles: 1) Project Manager (PM), 2) Requirements Engineer (RE), 3) Software Architect (SA), 4) Quality Engineer (QE), 5) Configuration Manager (CM), and 6) Sales Engineer (SE). Detailed descriptions of the roles are available in the project instructions~\cite[Sec 5.1]{github}.

\subsection{Intergroup Relations in an Ecosystem} \label{sec:relations}
\begin{figure}
\centering
\includegraphics[width=0.5\textwidth]{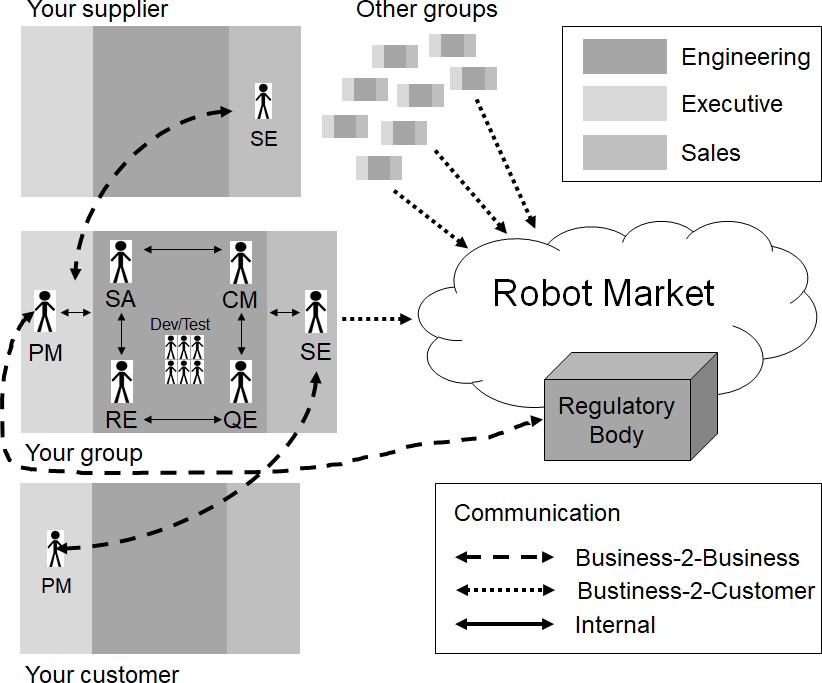}
\caption{The business ecosystem in the course. Groups interact with the supplier group's SE, the customer group's PM, the open Robot Market, and the Regulatory Body.}
\label{fig:ecosystem}
\end{figure}

Groups do not complete the projects in isolation, instead they act in a business ecosystem as illustrated in Fig.~\ref{fig:ecosystem}. During the engineering of the robot, each group will be part of two supplier-customer relationships. First, the sales engineer is responsible for promoting sales of the robot under development. At the end of the first development sprint, all groups take part in a closed bidding process during which the teachers, referred to as the ``Regulatory Body'', ensure that each group gets one supplier and one purchaser (see Section~\ref{sec:business}). The project manager of the purchasing group will sign a contract of sale, establishing a formal relationship. Subsequently, the sales engineer is the primary communication point for the purchasing group's project manager, i.e., a supplier-customer relationship involving feedback, feature requests, and negotiations. Second, analogously, the project manager is the customer in a supplier-customer relationship with another group.  Finally, groups practice one-way communication with the \textit{Robot Market}, i.e., the open market where groups offer robots after at the end of the project, further explained in Sections~\ref{sec:process} and~\ref{sec:business}. 

The software business aspects of the project involve a combination of bespoke SE and market-driven SE. All project groups offer their robot on a highly competitive market -- consisting of the other groups, each with a \texteuro 100 budget to invest in a robot team for the rumble. The monetization activity is split into two parts: 1) bespoke supplier-customer relationships and 2) selling robots on Robot Market, further described in Section~\ref{sec:business}.

All groups complement their bespoke robot with additional robots to build a team for the rumble during the Post release stage (cf. Fig.~\ref{fig:ecosystem}). The regulatory body presents rules that robot teams must adhere to, thus variations between course editions can easily be achieved. On the other hand, what makes a successful robot team depends heavily on what the other teams do -- thus the metagame~\cite{boluk_metagaming:_2017} anyway changes each time.

\subsection{Process Model with Three Sprints} \label{sec:process}
Most of the project weeks are devoted to the second phase in Fig.~\ref{fig:overview}, i.e., engineering, monetizing, and strategizing. 
During this phase, groups work in three two-week sprints with fixed deliverables.
Engineering requires more effort than the other three activities, thus we further organize it in the sub-activities \textit{specification}, \textit{construction}, and \textit{verification}. 
Since we teach iterative development, we stress that all sub-activities are involved in all sprints, but, each sprint will focus on one of the sub-activities. 

\begin{figure}
\centering
\includegraphics[width=0.50\textwidth]{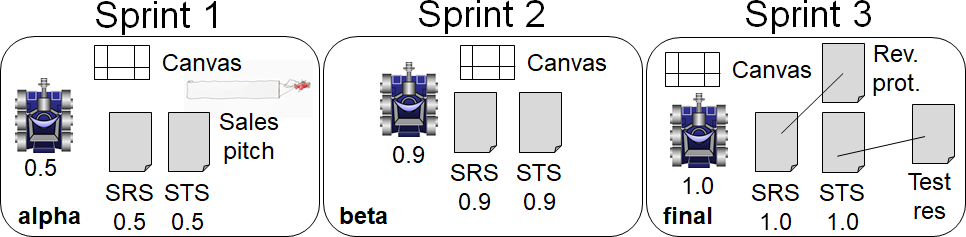}
\caption{The process model organized in three sprints with the corresponding deliverables.}
\label{fig:process}
\end{figure}

To ensure iterative development based on customer interaction, each group makes a release of their robot after each sprint, see Fig.~\ref{fig:process}. A release is defined as an executable robot and the latest version of the following artifacts: lean canvas (described in Section~\ref{sec:business}), Software Requirements Specification (SRS), and Software Test Specification (STS). During \textbf{Sprint 1}, students focus on on engineering (mostly specification) and monetizing, The goals are to do 1) feature scoping and 2) robot prototyping, and to 3) develop a sales pitch (described in Section~\ref{sec:business}). Groups prepare an alpha release (v0.5) -- a prototype. During \textbf{Sprint 2}, students focus on engineering (mostly construction) and strategizing. The goals are to 1) evolve the robot, 2) maintain the customer relationship, and 3) develop a strategy for the rumble. Groups prepare a beta release (v0.9) -- a minimum viable product. During \textbf{Sprint 3}, students focus on tying up all activities and preparing the final release (v1.0). The engineering focus is on verification. The goals are to 1) complete the robot development, 2) generate high-volume sales, and to 3) optimize the rumble strategy. Groups complement the final release with a test result protocol corresponding to the STS v1.0 and a review protocol for the final penultimate version of the SRS.


One might wonder why we do not teach an agile development method since this often is the preferred approach in industry. The primary reasons are a lack of teaching resources and the challenge to schedule sessions for two different engineering programs. To properly teach agile methods, there is a need for considerable teacher presence during the software development and management -- this is costly and hard to schedule for diverse student cohorts. Previous work presented a successful structure for teaching extreme programming~\cite{beck_extreme_2000} to large groups of students by designing courses that work in tandem~\cite{hedin_teaching_2005}, i.e., one course where a large group of junior students learns an agile method and one course where senior students learn to act as agile coaches. The proposed approach appears crafty and cost-effective, but it is not realistic in our case. Instead, we supervise student development closely during lab sessions, and particularly recommend a set of agile practices that we find valuable in our educational context: pair programming, continuous integration, code refactoring, test-driven development, and collective code ownership~\cite[Section 2.1.2]{github}.

\subsection{Software Business and Decision Making} \label{sec:business}
The business components in the course include niche markets, sales pitching, customer negotiations, and price setting. Groups document their business plan using a Lean Canvas~\cite{maurya_lean_2012} -- the canvas must be maintained to keep business and engineering aligned.

Robocode uses three categories of robots, which we use to create niche markets. By tailoring rules for the rumble, the Regulatory Body can incentivize niches. In our latest edition of the course, we stated that a robot team must consist of one leader robot, [0-4] normal robots, and any number of droids. We require that each robot is put to the market as one of the three categories, effectively creating niche markets for leaderbots and droids. Consequently, groups must decide whether to develop a normal robot or to target a niche. 

The bespoke SE is established through a ``Robot Fair'' after Sprint 1. At the fair, each group shows a 2-minute video pitch of the main features their planned robot will offer at the rumble~\cite[Sec 6]{github}. Groups are also welcome to share any other material for sales promotion. One day after the robot fair, all groups take part in a sealed bid auction by submitting a ``purchase array'' to the Regulatory Body consisting of one bid for each of the robots offered by others. The Regulatory Body then analyzes all bids and matches suppliers with the customers willing to pay the highest prices. Each group establishes two relationships of bespoke SE, one as a supplier and one as a customer~\cite[Sec 2.2.1]{github}. Sealed bid auctions have been used for fair allocations in SE education before~\cite{beck_fair_2008}, and we are satisfied with the results. 

Groups must balance the relation with the bespoke customer with the market-driven SE targeting the Robot Market. The same robot developed in the bespoke context will also be offered on the open market in the Post release phase (cf. Fig.~\ref{fig:overview}), i.e., groups need to practice customer negotiation while evolving the SRS since different variants of the robots are not allowed. If a customer's acceptance testing finds that the final release of the robot does not fulfill the SRS, they can file a business claim to the regulatory body to request a cost deduction. On the other hand, if the supplier group finds that their customer requests strange customization of features during development that might make the robot less attractive on the open market, important decision making must be made -- based on both technical and business considerations. This teaches students to tackle realistic uncertainty.

Finally, groups practice software price setting on the Robot Market. Robot Market is regulated, and groups are not allowed to price their robot below what the bespoke customer has paid. However, groups can offer robot bundles with discounts for buyers that purchase more than one robot~\cite[Sec 2.2.2]{github}. Since the robots are virtual, there are no traditional production costs involved. Groups must decide whether to market an expensive premium robot or to aim for high-volume sales through more aggressive pricing.

\section{Research Method} \label{sec:method}
This section presents the intervention we did between the two consecutive course editions, how we collected data to assess its effects, and the subsequent qualitative and quantitative analyses.

\subsection{Intervention -- Mandatory Lab Sessions} \label{sec:intervention}
We made one major change to the course before the 2019 edition, which motivated a minor revision of the roles in groups. First, the previously offered ``computer exercises'' were replaced by mandatory ``lab sessions'' -- but the number of sessions on the schedule and their content remained the same. To pass each lab session, students must report how they completed the tasks described in the lab instructions. Due to limited resources, we encourage students to work in pairs. Absent students are allowed to complete the labs on their own.

With mandatory lab sessions, all students effectively become developers and testers (cf. Fig.~\ref{fig:ecosystem}). To stress that students should not leave development or testing to individual students, we updated the 2018 roles accordingly. The role of development lead was renamed to software architect. The role of test lead was renamed to quality engineer. Finally, due to numerous student issues with git during the 2018 edition, we replaced the role of Robocode domain expert with a configuration manager expected to help the group with git.

\subsection{Data Collection and Analysis}
\begin{figure}
\centering
\includegraphics[width=0.45\textwidth]{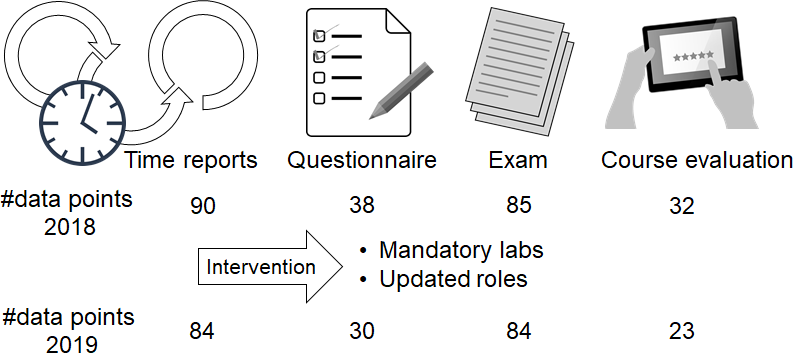}
\caption{Instruments used in the study. The figures show the number of collected data points.}
\label{fig:collection}
\end{figure}

Fig.~\ref{fig:collection} shows an overview of our measurement instruments. We collected data in analogous ways during the course editions in 2018 and 2019. To measure the effect of the intervention, we used an identical combination of quantitative and qualitative measurement instruments both in 2018 and 2019. \textbf{Time reports}: Each week, students reported the full hours spent on the project. 
\textbf{Questionnaire}: During the penultimate lecture, we did a 40 min summary of the course. For each activity in the course, students provided a fixed-scale rating (1-5) reflecting satisfaction along with free-text motivations and improvement suggestions.
\textbf{Exam}: The course concludes with a 26 h individual take-home essay exam. Both in 2018 and 2019, the final exam question was a retrospective analysis of work distribution during the project.
\textbf{Course evaluation}: After the exam, but before its grading, the students received the formal online course evaluation questionnaire with 26 questions and free-text fields.

The quantitative data from the time reports are presented using histograms with kernel density estimates~\cite{silverman_density_1998}. We conduct Levene's test ($\alpha$=0.05) to investigate equality of group variances between the total student efforts in 2018 and 2019. Since the fixed-scale ratings of the questionnaire are measured on an ordinal scale, we analyze them using median scores and interquartile ranges. The formal course evaluations are independently analyzed by a central function at the university, based on descriptive statistics and the Kruskal-Wallis test ($\alpha$=0.05) to investigate statistical differences between satisfied and dissatisfied students. 

The qualitative data from the exam questions are our richest source of information. We conducted thematic coding while grading the 2018 exams. According to Braun and Clarke, ``a theme captures something important about the data in relation to the research question, and represents some level of patterned response or meaning within the data set.''~\cite[p. 82]{braun_using_2006}. Codes emerged through an iterative process.

When coding the 2019 exam, we found that the 2018 codes could be applied without any modifications -- except one code related to the ```domain expert'' role that was replaced for 2019 (see Section~\ref{sec:intervention}. To support robustness, we report only codes that were used at least 10 times in total. Moreover, the responses were anonymized but the information about the role and whether the student followed the ICT or MGM program was kept. For both sub-questions in the retrospective exam question, we organize the answers into 1) phenomena \textit{experienced} by the students and 2) \textit{improvement proposals}. Some codes appear in both 1) and 2) since they represent phenomena that were experienced by some groups and wanted by others.

When quoting students in Section~\ref{sec:res}, we encode source information in brackets. We use the following format [ProgramYear-Role], e.g., [ICT18-PM] for an ICT student that followed the course in 2018 as a project manager. When quoting from the formal course evaluations, we omit the role as this information is missing.

\section{Results and Discussion} \label{sec:res}
This section presents our results and answers the three RQs.

\subsection{RQ1: Individual Work Balance Over Time}
We hypothesized that if students receive increased hands-on development training through mandatory lab sessions, they would be able to contribute to the project earlier and thus balance their weekly efforts more evenly. Table~\ref{tab:weekly} shows students' reported hours before and after the intervention. In 2018, students reported roughly 6h for the first week, followed by two weeks of holidays, and then weeks 5 and 7 stand out as more project-intensive. In 2019, we indeed notice increased project effort the first week, but also a final week of evident ``crunching''~\cite{borg_video_2019}. We also find that the less experienced ICT students, with limited programming skills, do not make up for this by working harder than the MGM students -- even less after the intervention. It is possible that the ICT students were so occupied with completing the lab assignments that they did not have time for the project.

\begin{footnotesize}
\begin{table*}[]
\caption{Reported student effort per week. Average hours with standard deviations in parentheses. Note that due to how holidays fell, denoted in italic font, the students had an additional calendar week to complete the project in 2018.}
\begin{center}
\begin{tabular}{l|ccccccc|c}
Year & w1         & w2         & w3        & w4        & w5         & w6          & w7        & Total \\ \hline
2018 & 5.9 (3.9)  & \textit{4.1 (4.4)} & \textit{7.1 (5.5)} & 8.3 (5.2)  & 11.5 (6.4)  & 7.6 (6.8) & 14.1 (8.4) & 58.6 (22.6) \\
ICT  & 6.0 (3.4)  & \textit{2.6 (3.6)} & \textit{5.3 (5.1)} & 7.8 (6.2)  & 10.6 (5.5)  & 8.7 (7.5) & 14.4 (8.1) & 55.6 (23.4) \\
MGM  & 5.8 (4.3)  & \textit{5.2 (4.6)} & \textit{8.5 (5.4)} & 8.6 (4.3)  & 12.2 (6.9)  & 6.7 (6.1) & 13.9 (8.8) & 60.9 (21.8) \\ \hline
2019 & 10.7 (4.4) & 9.3 (4.4)  & 8.4 (4.8) & \textit{1.4 (1.7)} & \textit{5.8 (8.5)}  & 26.3 (12.6) &               & 61.9 (24.5) \\
ICT  & 9.4 (4.2)  & 8.0 (4.4)  & 8.0 (6.0) & \textit{1.3 (1.7)} & \textit{5.4 (10.5)} & 21.9 (14.2) &             & 53.9 (29.7) \\
MGM  & 12.0 (4.3) & 10.7 (4.0) & 8.8 (3.3) & \textit{1.5 (1.7)} & \textit{6.2 (6.0)}  & 30.7 (9.0)  &          & 69.9 (14.2)
\end{tabular}
\end{center}
\label{tab:weekly}
\end{table*}
\end{footnotesize}

The left component in Table~\ref{tab:codes} shows the results from coding the retrospective exam question related to individual work balance. Looking at the experienced codes, we find that \textit{Deadline crunching} was the most frequent in 2019 -- discussed by 82\% of the 2019 students compared to 51\% in 2018, i.e., substantially more often (+60\%) after the intervention. This finding is supported by the time reports presented in Table~\ref{tab:weekly}. We believe that the increased focus on completing lab assignments during the course put the project work on hold in some groups, thus students faced stress in the end. 

\begin{table*}[]
\caption{Codes for the exam question and their distributions. A heading + or - indicates positive vs. negative experiences.}
\includegraphics[width=1\textwidth]{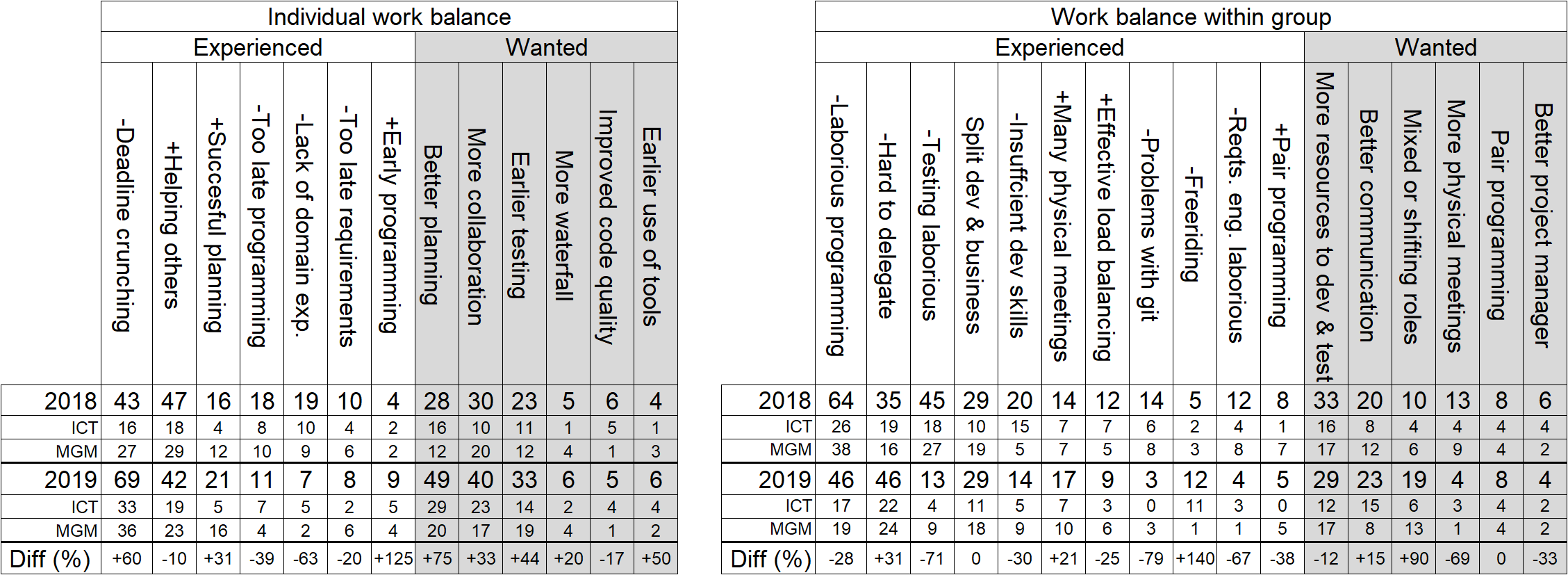}
\label{tab:codes}
\end{table*}

Before the intervention, the most frequent code was \textit{Helping others}. While we do observe a decrease after the intervention (-10\%), we do not believe students were less helpful in 2019. A possible explanation is that since all students now received basic development training, the collaboration around the source code became implicit, thus fewer students mentioned it in the retrospectives. Regarding the programming, we find two improvements after the intervention. First, \textit{Too late programming} was reported less frequently in 2019 (-39\%) and \textit{Early programming} was reported more often (+125\%). Both changes follow naturally from introducing mandatory lab sessions from the first week and the effect matches our goals with the intervention. Another positive effect was that the technical lab assignments helped students to get involved in the domain, i.e., Robocode. Fewer students (-63\%) reported \textit{Lack of domain experience} as an issue in 2019.

Considering improvement proposals by the students, i.e., the wanted codes, we observe three dominant themes: 1) \textit{Better planning}, 2) \textit{More collaboration}, and 3) \textit{Earlier testing} -- all three increased after the intervention. Thus, it appears that that mandatory lab sessions increased the need for planning and collaboration. We believe that providing basic development skills to all students enabled better work distribution, but the students realized that fundamental planning (+75\%) and collaboration (+33\%) were essential to make it work. More students also suggested earlier testing (+44\%), which again could be explained by the students' increased practical skills in what software testing involves.

Despite the confounding factor of fewer calendar weeks in 2019, we conclude that \textit{mandatory lab sessions did not lead to students distributing the work more evenly over the weeks}. Unfortunately, the phenomenon of deadline crunching was increasingly present after the intervention. However, student procrastination is a well-known problem with no quick fixes.

\subsection{RQ2: Work Balance Within Group}
\begin{figure}
\centering
\includegraphics[width=0.5\textwidth]{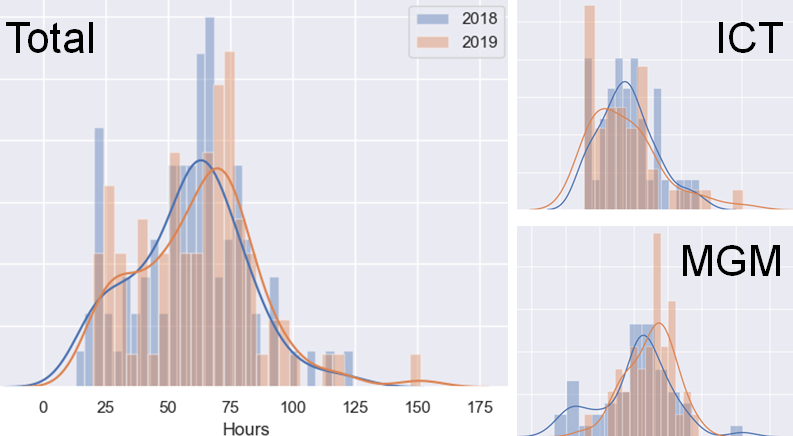}
\caption{Distribution of students' total effort.}
\label{fig:time}
\end{figure}

Fig.~\ref{fig:time} shows the distribution of total reported time per student. The (Total) plots for 2018 and 2019 look highly similar and Levene's test certainly does not reject the null hypothesis of equal variances (p=0.51). Separate analyses of the ICT and MGM students display the same similarity. From the perspective of fair workload sharing, the time reports show no effect of the intervention.

\begin{figure}
\centering
\includegraphics[width=0.5\textwidth]{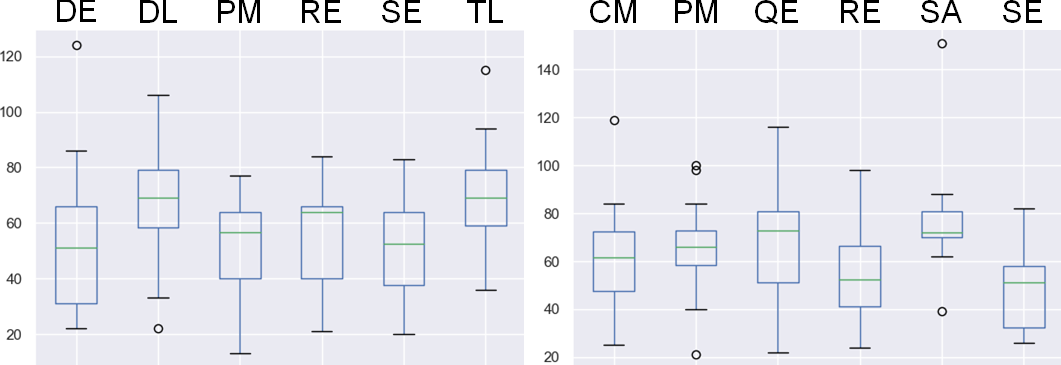}
\caption{Distribution of student effort per role.}
\label{fig:roles}
\end{figure}

Fig.~\ref{fig:roles} shows how much time different roles reported. In 2018, the Development Lead (DL) and the Test Lead (TL) were overburdened with work -- this was the primary motivation for the intervention. However, we again found that the corresponding roles, i.e., Software Architect (SA) and Quality Engineer (QE), reported more hours than other roles. The median values for the comparable roles look similar over the years, except for the Requirements Engineer (RE) whose median hours decreased. A possible explanation is that students with less interest in programming were inclined to be REs. However, with mandatory lab sessions, they had to program anyway -- if they struggled with the programming, perhaps they had less time to work on requirements.  

The right component in Table~\ref{tab:codes} shows the results from coding the retrospective exam question related to work balance within the group. Starting with the experienced codes, we find that the students found the project to be programming heavy. After the intervention, however, \textit{Laborious programming} was mentioned less frequently (-28\%). The second most common code, i.e., \textit{Hard to delegate}, mainly captures the challenge of distributing programming tasks within groups -- a reflection that increased in 2019 (+31\%). The highlighted delegation challenge is in line with the students' increased calls for planning and collaboration reported for RQ1. Furthermore, the decreased frequency of the code \textit{Testing laborious} (-71\%) suggests that mandatory labs helped to balance the testing activity in the groups.

We found mixed results for some of the codes related to collaboration around the development tasks. As we hoped, we found a decrease in \textit{Insufficient dev. skills} (-30\%) and fewer \textit{Problems with git} (-79\%). On the other hand, making the lab sessions mandatory for all students had no effect on the (discouraged) tendency to \textit{Split dev \& business}, i.e., separating source code-oriented tasks (e.g., programming and testing) and business and user-oriented tasks (e.g., sales promotion and requirements engineering). Moreover, we observe an increase in reported \textit{Freeriding} after the intervention. We believe the increased collaboration stimulated by mandatory lab sessions made students more aware of non-performing group members.

Looking at the students' solution proposals to better balance the workload within groups, we find that the most common suggestion is to allocate \textit{More resources to dev \& test}. There is little difference after the intervention, suggesting that simply increasing the minimum set of dev/test skills provided by the mandatory labs is not enough to involve more students in the source code. Several retrospectives discussed motivation and personal interests. Some students enjoy programming and take an active role in the source code repository. However, another set of students would like to more actively contribute source code, but refrain as they feel unsure about their code quality.

The second most frequent solution proposal is \textit{Better communication}, both before and after the intervention. Finally, we note that \textit{Mixed or shifting roles} increased (+90\%) after the intervention. This was an unexpected finding, as the intervention was supposed to effectively mix roles by turning all group members into developers and testers. Again, the results show that more is required to successfully bring all students into the projects' source code.

We acknowledge that \textit{mandatory lab sessions did not reduce variance in the total effort per student}. Roles with source code responsibilities work more and find it hard to delegate. This suggests that we must explicitly teach how implementation tasks can be distributed.

\subsection{RQ3: Reception by Students}
We analyze the data from the interim questionnaire to assess whether the intervention had an impact on student satisfaction during the course. The responses from the questionnaires of 2018 and 2019 are comparable with response rates of 44.7\% and 35.7\%, respectively. Roughly the same number of ICT and MGM students answered both years and the distribution of roles among the respondents is similar. 

Table~\ref{tab:questionnaire} shows the results from the questionnaires before and after the intervention. The results suggest that students are more satisfied with all four lab sessions after the intervention. The unchanged exercise sessions can be seen as a baseline. However, the free-text answers from both 2018 and 2019 reveal that several students struggle with the labs as indicated by ``generally, additional teaching assistants are needed'' [MGM18-RE], ``challenging and a lot of issues with all labs'' [ICT18-PM], and ``extra sessions are needed with hands-on support by teaching assistants -- VERY hard to get started'' [ICT19-QE].

Labs 2 and 3 are more technically challenging and thus the answers are reflected by the differences between ICT and MGM students. A major difference can be seen in Table~\ref{tab:questionnaire} for Lab 3 in 2018, with considerably more satisfied MGM students -- but the difference disappeared after the intervention. Many students explain that Lab 3 was difficult, and report issues with RobotTestbed. While not shown in the ratings, the biggest contrast between ICT and MGM is shown in the free-text answers of Lab 2. ICT students have never seen unit testing before (e.g., ``too much information to digest and I didn't really get what it [unit testing] is good for'' [ICT18-SE]) and MGM students have worked with JUnit in a previous course (e.g., ``fairly easy lab since I've done it before, but it helped me structure my unit tests for the project'' [ICT19-PM]). In the next edition of the course, we will allocate more time and assistants to labs 2 and 3.

Looking at the specific skills in Table~\ref{tab:questionnaire}, we find that several figures are higher after the intervention. While there might be important confounding factors at play, further discussed in Section~\ref{sec:threats}, mandatory lab sessions have a direct influence on working in an IDE, programming, and testing. After the intervention, students are more satisfied with how they developed IDE and testing skills in the course. On the other hand, the impact on the development of programming skills is less clear -- ICT students report lower satisfaction, but the MGM students are very happy. Finally, there is a striking difference in how satisfied the ICT and MGM students are with the learning process of working in a team. After the intervention, the MGM students unanimously report the highest rating, but the ICT students are dissatisfied. The free-text answers indicate that several groups with ICT students did not perform as a team, e.g., ``I probably didn't contribute a single line of code, I didn't test anything, I didn't use git'' [ICT19-RE] and ``two of us did everything since the others were unreliable or lacked programming skills'' [ICT19-PM].

\begin{footnotesize}
\begin{table*}[]
\caption{Median scores (1--5) from the questionnaire with inter-quartile range (IQR) in parentheses. Satisfaction of labs and exercises to the left and specific skills to the right (RE=reqts. eng., Bus.=business perspectives, IDE=working in an IDE, Prog.=programming, Test=testing, CM=config. mgmt., Team=team work). Gray cell = increase, black cell = decrease}
\begin{center}
\includegraphics[width=0.951\textwidth]{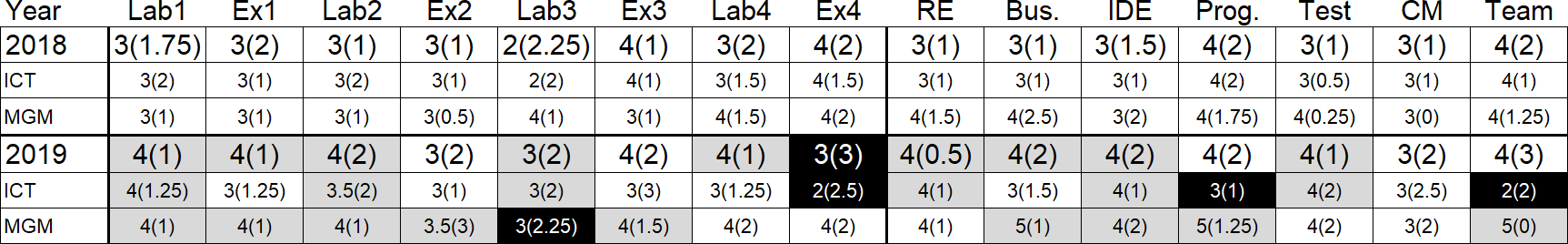}
\end{center}
\label{tab:questionnaire}
\end{table*}
\end{footnotesize}

\begin{figure}
\centering
\includegraphics[width=0.4\textwidth]{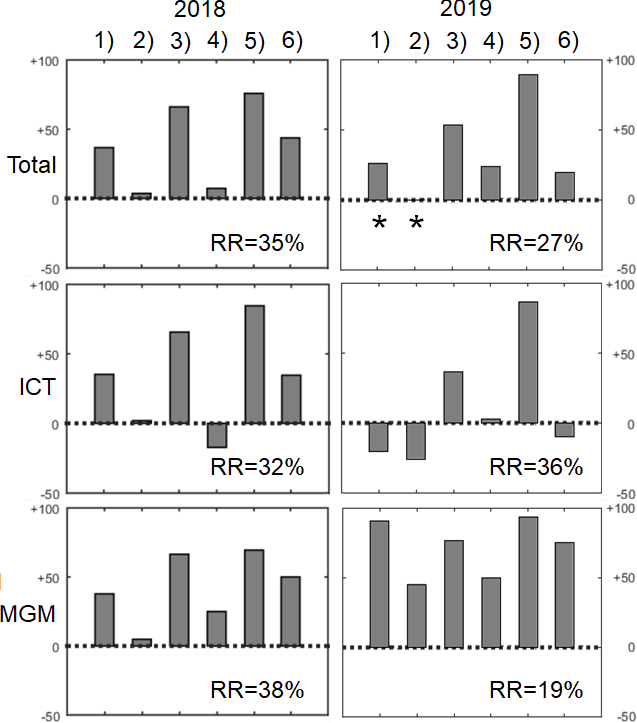}
\caption{Summary of the course evaluations. The two stars indicate statistically significant differences between satisfied and dissatisfied students.}
\label{fig:ceq}
\end{figure}

Finally, we assess the publicly available course evaluations
analyzed centrally by the university. Fig.~\ref{fig:ceq} presents the summary of the 26 questions, grouped into six categories with scores spanning between -100 och +100. From left to right, the bars depict the categories: 1) Good Teaching, 2) Clear Goals, 3) Appropriate Assessment, 4) Appropriate Workload, 5) Importance for Education, and 6) Overall Satisfaction. As usual, the Response Rates (RR) are low with the biggest difference between MGM students of 2018 and 2019.

After the intervention, we observe significant differences between satisfied and dissatisfied students related to Good Teaching and Clear Goals. Dissatisfied students did not appreciate the teaching and found the goals unclear. Fig.~\ref{fig:ceq} clearly shows that the ICT students are critical while the MGM students give the teaching top ratings. This finding is in contrast to the questionnaire that showed increased lab satisfaction among ICT students. Still, we believe that the mandatory lab is the reason behind the difference. The student cohorts of 2018 and 2019 are similar, but before the intervention, weak programmers could either watch their friends solve the lab assignments or entirely skip the sessions. In 2019, also weak programmers (often pairs of weak programmers) had to complete tasks that were too difficult -- resulting in a subset of dissatisfied students.

This interpretation is supported by statements such as: ``The labs often resulted in frustrated chaos since they were poorly planned. The preparation tasks were rarely sufficient to help us get started with the labs.'' [ICT19] and ``Don't give students such a huge workload. It would have been good to complete labs before the corresponding project tasks, but most did them afterward. It was catastrophic when we had enormous tasks that nobody understood.'' [ICT19]

Also for the other categories, we find that the intervention impacted the ICT and MGM students differently. For the ICT students, we find decreased scores for all categories but Appropriate Workload and Importance for Education. For MGM students, on the other hand, all six categories got increased ratings. The satisfaction of the MGM students becomes even more evident in statements such as: ``The first course that actually feels related to working life. So much fun!'' [MGM19], ``The engaged teachers are great, the course design is inspiring, and there is plenty of hands-on work rather than just studying the theory'' [MGM19], and ``The best course I've ever taken. Praise to the lecturer whose passion for the subject permeated the period. The course is very applied /\ldots/ programming is fun and creative.'' [MGM19] 

We conclude that \textit{the students' perceptions of mandatory labs span the entire spectrum} and their \textit{programming background is the main variation factor}. We designed the intervention to support novices but the content largely went over their heads. However, as course evaluations can be weak predictors of learning outcomes~\cite{carrell_does_2010}, it is possible that the dissatisfied ICT students indeed learned valuable SE skills.

\section{Threats to Validity} \label{sec:threats}
As elaborated by Hutchinson~\cite{hutchinson_evaluating_1999}, there are fundamental difficulties when evaluating educational interventions. What works, for whom, in what context, and at what cost? Our findings are subject to important threats to internal validity, as many confounding factors are inevitably at play. Making causal claims regarding specific interventions based on a single study is questionable -- thus our conclusions are defensive. While we designed the study intending to fix as many variables as possible, some could not be controlled. Important changes between the course editions that might have influenced the intervention include: 1) some teaching assistants were replaced, 2) remaining teachers might have done a better job the second year, and 3) the holidays fell differently, with a big irregular student festival in town in 2018.

We also highlight some threats to construct validity and external validity, respectively. Students wrote retrospectives as part of a graded exam, thus they might have provided answers with the goal to please the teachers, i.e., a type of response bias~\cite{berk_beyond_2004}. When self-reporting hours for the time reports, students might be subject to conformance bias~\cite{padalia_conformity_2014} and report figures matching the other group members. Finally, educational interventions are highly context-dependent, thus analytical generalization is required to extrapolate from our work~\cite{kvale_interviews:_1996}. To support this, we provide a rich case description.


\section{Conclusion} \label{sec:conc}
Project courses with group work are important in SE education but bring challenges when teaching diverse student cohorts. In a course combining software and business development in groups of six students, we noticed a considerable gap in programming skills, resulting in students completing the course without ever working on the source code level. To close this gap, we designed an intervention to make lab sessions mandatory.

We found that the intervention was very well received among mature students, but not popular with students weak at programming. While the goal was to help novices learn fundamental development skills to enable more collaboration on the source code level, we instead found that weak programmers now struggled with the lab assignments and instead had less time to contribute to the projects at all. Moreover, the intervention did not result in more balanced work distribution neither within the groups nor for individuals over time.

We conclude that more than mandatory labs are needed to balance the programming workload among differently skilled students. However, we still believe that the intervention is a step in the right direction. For the 2020 edition of the course, we will extend the time of the lab sessions and increase the number of teaching assistants.

\bibliographystyle{IEEEtran}
\bibliography{seet}

\end{document}